\documentclass{PoS}

\def\solm{M$_{\odot}\,$}

\title{Galaxy Formation: Where Do We Stand?}

\ShortTitle{Galaxy Formation: Where Do We Stand?}

\author{\speaker{Christopher J. Conselice} \\
        Centre for Astronomy and Particle Theory\\
        University of Nottingham, United Kingdom\\
        E-mail: \email{conselice@nottingham.ac.uk}}

\abstract{This paper presents a review of the topic of galaxy formation and
evolution, focusing on basic features of galaxies, and how these observables 
reveal how galaxies and their stars assemble over cosmic time.  I give an
overview of the observed properties of galaxies in the nearby universe and
for those at higher redshifts up to $z \sim 10$.  This includes a discussion of
the major processes in which galaxies assemble and how we can now observe
these - including the merger history of galaxies,
the gas accretion and star formation rates.  I show that
for the most massive galaxies mergers and accretion are about
equally important in the galaxy formation process between $z = 1-3$, while 
this likely differs for lower mass systems.  I also discuss the mass
differential evolution for galaxies, as well as how environment can affect
galaxy evolution, although mass is the primary criteria for driving 
evolution.    I also discuss how we are beginning to measure the 
dark matter content of galaxies at different epochs as
measured through kinematics and clustering. Finally, I review how
 observables of galaxies, and the
observed galaxy formation process, compares with predictions from
simulations of galaxy formation, finding significant discrepancies in the 
abundances of massive
galaxies and the merger history.  I conclude by examining prospects for
the future using JWST, Euclid, SKA, and the ELTs in addressing outstanding issues.  }

\FullConference{VIII International Workshop on the Dark Side of the Universe,\\
		June 10-15, 2012\\
		Rio de Janeiro, Brazil}

\begin{document}

\section{Introduction}

Anyone trying to understand the formation of the universe must consider galaxies.  Often used as a tracer population of the underlying matter, it is becoming clear that the history and physics of the formation of galaxies is a critical aspect for obtaining a full picture of the evolution of the universe.  Yet we are really just starting to understand this, and uncertainties in our measurements of evolution remain large, although significant progress has been made in outlining the basic problem.  

What is better understood is the observations of galaxies, which ultimately should reveal how the galaxies themselves have formed and evolved. In the local universe we have a solid understanding of the galaxy population based on large surveys such as SDSS and GAMA (e.g.,
Driver et al. 2011). Galaxies which are elliptical/passive along with spirals that have active star formation dominate the 
local population.  The stellar and luminosity functions of nearby galaxies are also well measured (e.g., Loveday et al. 2012) and we 
now understand when the stars in these galaxies were formed based on detailed stellar population analyses.

What we do not yet understand is how or when these galaxies assembled.  There has however been significant progress on addressing some of the issues related to this in the last 10-15 years.  For example, the first measurements of galaxy properties at high redshift showed that the star formation rate is larger at higher redshifts  than at lower redshifts (e.g., Madau et al. 1998), but this only reveals when stars form, not necessarily when galaxies assemble.  We have some idea of this through observing the merger history of galaxies (e.g., Le Fevre et al. 2000; Conselice et al. 2003, 2008; Hammer et al. 2009; Lotz et al. 2008; 2011; Bluck et al. 2009, 2012; Man et al. 2012; Lopez-Sanjuan et al. 2012), and other modes such as gas accretion are now also being measured (e.g., Conselice et al. 2012).  It is also important to realise that the questions of when the stars form in galaxies we see today, and when the galaxies themselves form, are distinct.  It is also important to remember that these issues often must be addressed separately.  

Traditionally the approach taken towards understanding the physics behind galaxy formation is to use galaxy observables, such as luminosity/mass functions, relations between observables, such as velocities and luminosities to test models.  These models, starting with simple collapse ones (Eggen et al. 1962) implement basic physics, and subsequently predict the observables.  This is now a large industry, and theory has revealed many clues towards understanding how galaxy formation has occurred, although as we discuss later (\S 5) there are still significant problems with theory matching galaxy observables.

\begin{figure}[b]
\begin{center}
 \includegraphics[width=5.5in]{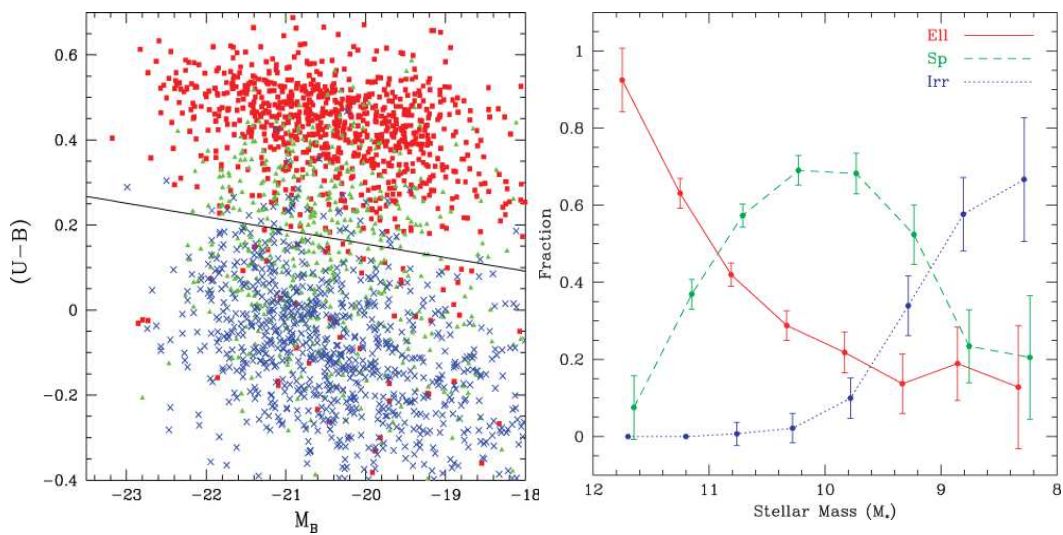} 
 \caption{Plots showing basic features of the nearby galaxy population as
a function of stellar mass.  The left panel shows the colour-mass
diagram for nearby systems, showing a clear differential between red
galaxies on the red sequence and blue galaxies in the blue cloud. This
correlates very strongly at $z = 0$ with galaxy morphology with ellipticals
found in the red sequence (red points) and spirals/mergers (blue points) 
in the blue cloud.  The
right panel shows the distribution of morphology as a function of stellar mass in
the nearby universe,
with a smooth transition from early to late types at lower masses (Conselice 2006). }
   \label{fig1}
\end{center}
\end{figure}

Overall, in this review I place galaxy studies into three different classes: (1) Observables (i.e., measured or derived directly from telescopes) - this includes galaxy masses, luminosities, internal velocities, sizes, morphologies, etc.  (2) Measuring galaxy history, i.e., how does the mass assembly/star formation, merging, and evolution in scaling relationship occur?  (3) Finally, physics - what are the mechanisms driving the history of the observed assembly? An important side question is whether we have identified all of these processes.   Within this framework the observables are first needed, then the history of the galaxy formation can be derived, which influences and leads into our understanding of the physics.    Each of these three are active areas of research and can and often help guide the development of the other two.  

This review will describe the observations of galaxies up to redshifts of $z \sim 10$, and how we are now able to measure the history of galaxy formation in detail up to at least $z = 3$, but only currently for the most massive systems with log M$_{*} > 11$.  For these systems we are obtaining a good idea of the formation mechanisms of galaxies.   We demonstrate this through measuring the merger history of these galaxies, and the amount of cold gas accreted from the intergalactic medium at redshifts $1 < z < 3$.  We show that these two processes are about equal in importance, and that minor mergers are as important as major mergers in forming galaxies.   We also discuss how well theory is able to reproduce some of these properties, and discuss significant issues that still needing addressing.  We use a standard cosmology of H$_{0} = 70$ km s$^{-1}$ Mpc$^{-1}$, and $\Omega_{\rm m} = 1 - \Omega_{\lambda}$ = 0.3 throughout.

\section{Observations}

\subsection{Nearby Galaxies}

Nearby galaxies at $z < 0.3$ have been studied in detail since the 1920s, and in many ways we know the most about galaxy properties and observables from examining nearby systems.  What we know is that 75\% of galaxies brighter than M$_{\rm B} = -20$ are spiral or disk in morphology, with 22\% SO/elliptical, and the remaining 2\%  are peculiar/irregular (Conselice 2006). Being nearby and thus relatively easy to study, many properties of nearby galaxies have been measured in detail, including their stellar mass and luminosity functions (e.g., Loveday et al. 2012; Taylor et al. 2011), as well as their detailed surface brightness distributions and internal kinematics (e.g., Cappellari et al. 2011; 
Kelvin et al. 2012).  

Perhaps the most useful observation that can be performed on nearby galaxies in terms of their formation histories is to study in detail their stellar populations through absorption lines (e.g., Trager et al. 2000; Thomas et al. 2005).  These studies have found that the formation history of the stars in galaxies depends strongly upon  their stellar mass, and less so on their environment.    The most massive systems are dominated by old stars, such that single stellar population ages of their stellar masses are older than 5 Gyr  (Trager et al. 2000).      By comparing absorption line strengths to models, it can be shown that the most massive
galaxies also contain the most $\alpha$ enriched stellar populations, 
demonstrating a quick formation for the stellar populations within these massive galaxies (e.g.,  Thomas et al. 2005).   
The stars in lower mass systems form later, although this may not be the case in the densest environments.  There is furthermore a lack of strong environmental effects, such that galaxies in the most dense areas appear to have similar ages as those in lower densities.  The effects of environment appear to be less important, and progressively so at higher redshifts (e.g., Gr\"utzbauch et al. 2011a,b; \S 4.2).

This is also seen in the distribution of morphology and star formation 
in the local universe as a function of stellar mass, whereby the lowest mass galaxies are the most likely to 
have a spiral or irregular morphology, and higher star formation rates and bluer
colours (e.g., Baldry et al. 2006; Conselice 2006; see Fig. 1). There is in fact a very
strong correlation such that the value of a galaxy's stellar mass, averaged over all 
environments, is a strong predictor for the morphology and star formation histories of 
individual galaxies.  

In terms of galaxy evolution, another important insight that local galaxies provides is a nearby benchmark by which we can gauge how galaxy properties have changed as a function of redshift.  In this review, I focus on the evolution of galaxy masses,  morphologies, and kinematics. I do not provide a detailed study of other features of nearby galaxies here, such as how galaxy clustering or environment correlates with the properties or evolution of nearby galaxies, although see e.g., Kauffmann et al. (2004) for a discussion of this.

\begin{figure}[b]
 \vspace*{-1.5 cm}
\begin{center}
\hspace*{-1.1cm} \includegraphics[width=6.6in]{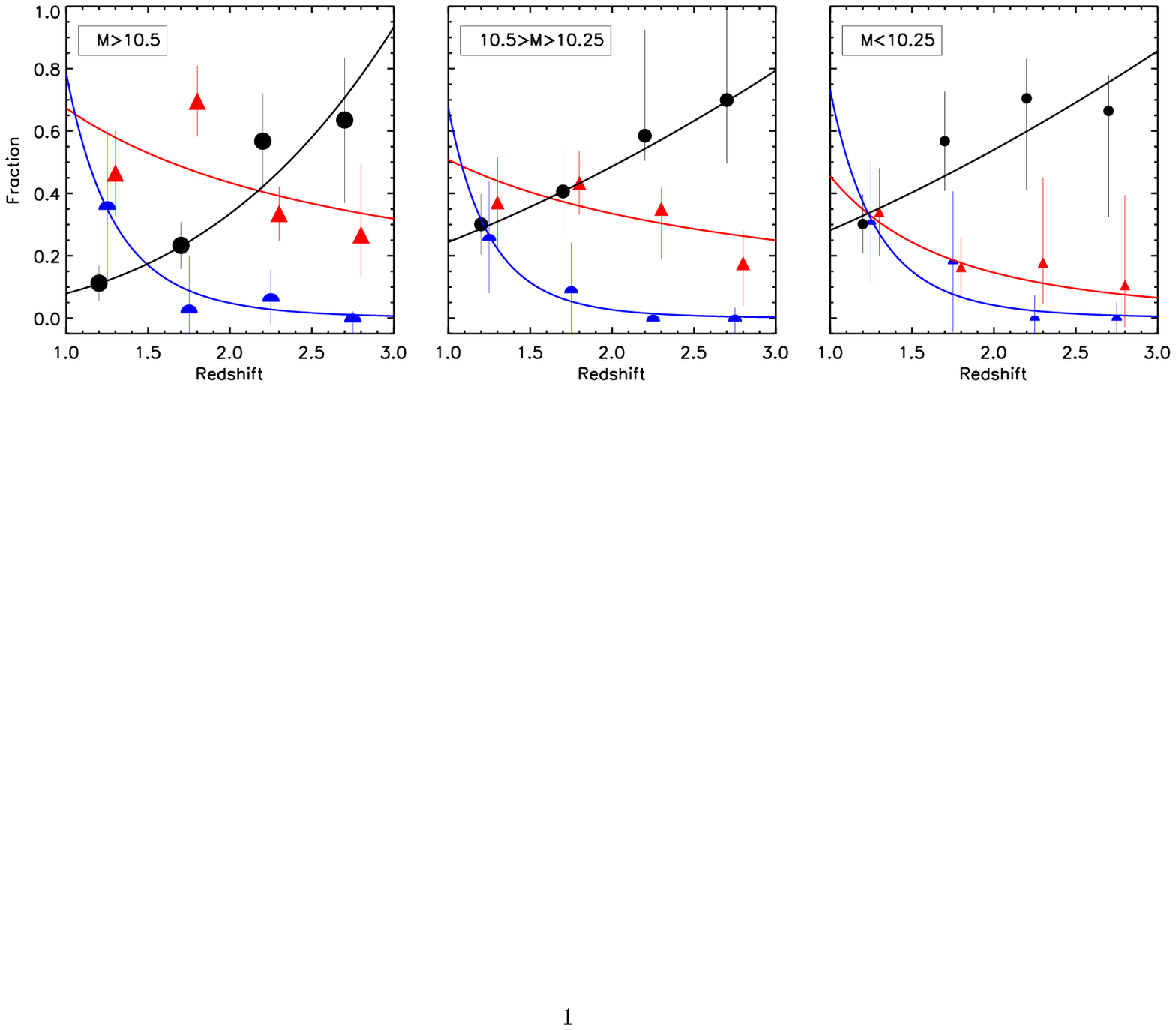} 
\vspace*{5.0 cm}
 \caption{Morphological evolution of the galaxy population from Mortlock et al. (2013). The
black lines are for galaxies classified as peculiar (as in Conselice et al. 2011b), and
the red lines are for galaxies classified as ellipticals and the blue lines
those as disks.  These have been corrected for redshift effects by determining
how structure changes due to redshift effects as detailed in simulations (see
Mortlock et al. 2013).  There is a clear evolution in structure, such that the
most massive galaxies form `normal' Hubble types before lower mass systems.   }
   \label{fig1}
\end{center}
\end{figure}

\subsection{Distant Galaxies: The Universe at $z > 1$}

The epoch of high redshift studies of galaxies began in earnest due to deep
Hubble Space Telescope observations (e.g., Ferguson et al. 2000), and with the
first 8-10m telescope observations of Lyman-break galaxies (LBGs) (Steidel et al.
1996), both quickly leading to the first measurements of the star formation
history of galaxies (e.g., Madau et al. 1996).  Since that time, observations
with the HST have pushed observations of LBGs up to $z = 8$ and perhaps 
beyond (Yan et al. 2012) up to $z = 10$ (e.g., Coe et al. 2012).  While
we are still detecting galaxies at progressively higher redshifts, and have
unlikely seen the first galaxies yet, we are able to make  observations
of these systems, during the $\sim$13 Gyr time period  when most galaxy assembly has 
occurred.    

The observables that are most commonly investigated for galaxies beyond the
local universe are similar to what we can measure for nearby galaxies, including: masses, 
luminosities, morphologies, sizes, kinematics and clustering, as 
well as derived quantities such as star formation rates and gas masses.  Observationally
is makes sense to divide high redshift galaxies into four epochs, largely for
observationally reasons. These are: $z < 1$, $1 < z < 3$, $3 < z < 6$ and
$z > 6$.  Each of these epochs is studied in slightly different ways, and obviously
we know progressively less about the more distant and earlier epochs.  

The universe up to $z \sim 1$ (at half its current age) in many ways `looks'
similar to the universe today.  It is still uncertain what fraction of
the stellar mass was assembled by $z = 1$, but it is likely more than
half (e.g., Mortlock et al. 2011).  The morphologies of $z = 1$ galaxies
are also very similar to those that we see in the local universe with
a similar fraction of different types (e.g., Conselice et al. 2005a),
however these galaxies have more of a clumpy structure overlayed on their
primary Hubble types (spirals, ellipticals) (e.g, Elmegreen et al. 2007). 

At higher redshifts, particularly at $z > 2$ the universe of galaxies
is quite different from today. In general, galaxies are bluer, smaller,
more asymmetric, and contain higher star formation rates.  Attempts to
understand the evolution of this includes:  measuring luminosity and
mass functions (e.g., Bouwens et al. 2011); evolution of stellar populations
and colours (e.g., Finkelstein et al. 2012); morphological evolution, including
sizes and surface brightness profile evolution (e.g., Conselice
et al. 2005a; 2008; 2011; Buitrago et al. 2011; Weinzirl et al. 2012); 
clustering 
evolution (e.g., Foucaud et al. 2011) and star formation evolution
(e.g., Bouwens et al. 2010).  
 Also, $z > 1$ systems have smaller radii at a given
mass than local galaxies at the same stellar mass (e.g., Trujillo et al. 2007; 
Buitrago et al. 2008),
suggesting some evolution throughout time to increase the sizes
of galaxies by some still uncertain process (Ownsworth et al. 2012).  
 See these papers for details of these various
issues.

As one example, Figure~2 shows the morphological evolution for galaxies from the CANDELS
survey (Mortlock et al. 2013 in prep), demonstrating how there is both a 
redshift and stellar mass dependence on morphology.  As can
be seen at $z = 3$ the morphological fraction is dominated by galaxies
that appear peculiar in appearance (the black circles and lines), while
at lower redshifts, towards $z = 1$ the classical Hubble type galaxies -
those identifiable as disks or ellipticals start to become the dominant
population (e.g., Buitrago et al. 2011).  The other interesting feature of 
this evolution is that there
is a mass dependence such that this `transition' from peculiar to
normal galaxy is at higher redshifts for higher mass galaxies. This
transition redshift is $z = 2.22\pm0.87$ for log M$_{*} > 10.5$ galaxies,
$z = 1.75\pm0.76$ for $10.25 <$ log M$_{*} < 10.5$ galaxies, and 
$z = 1.73\pm0.56$ for log M$_{*} < 10.25$ systems.  

There is also strong star formation evolution from $z = 0$ to $z = 1$, 
such that the total star formation rate density averaged
over all galaxies at $z \sim 1$ is a few times higher per co-moving 
volume element than in today's universe (e.g., Bouwens et al. 2010).   
However, the star formation rate peaks and is roughly constant at a 
given stellar mass
between $1 < z < 3$ (Fig. 3), such that a typical massive galaxy will double
its stellar mass due to this star formation over the epoch $1 < z < 3$, and
lower mass galaxies grow even larger.   Later we investigate the relative
role of this star formation in growing the stellar masses of galaxies, and
what this reveals about their formation (\S 3).    At even higher redshifts
the total star formation density declines at $z > 3$ up to $z \sim 7$
(Bouwens et al. 2010) although the specific star formation rate for
the most massive galaxies is relatively constant (Stark et al. 2009), 
suggesting some star formation regulation process might be at work.

Perhaps the most basic way to address when galaxies form is however
to compare the stellar mass
distribution of distant galaxies to that of galaxies in the nearby universe.
 This is particularly the case for
the most massive systems, as these form before low mass galaxies (e.g., Bundy
et al. 2006; Mortlock et al. 2011), and are the easiest galaxies to simulate
in computers and are thus an invaluable test of galaxy formation models (e.g.,
Conselice et al. 2011b). 
Massive galaxies have been studied in detail using various surveys from the 
Hubble Space
Telescope within deep pointed surveys (Mortlock et al. 2011), as well 
wider area surveys using imaging from telescopes such as UKIRT within 
the UKIDSS Ultra Deep Survey (Hartley et al. 2010).

\begin{figure}[b]
\begin{center}
 \includegraphics[width=5.2in]{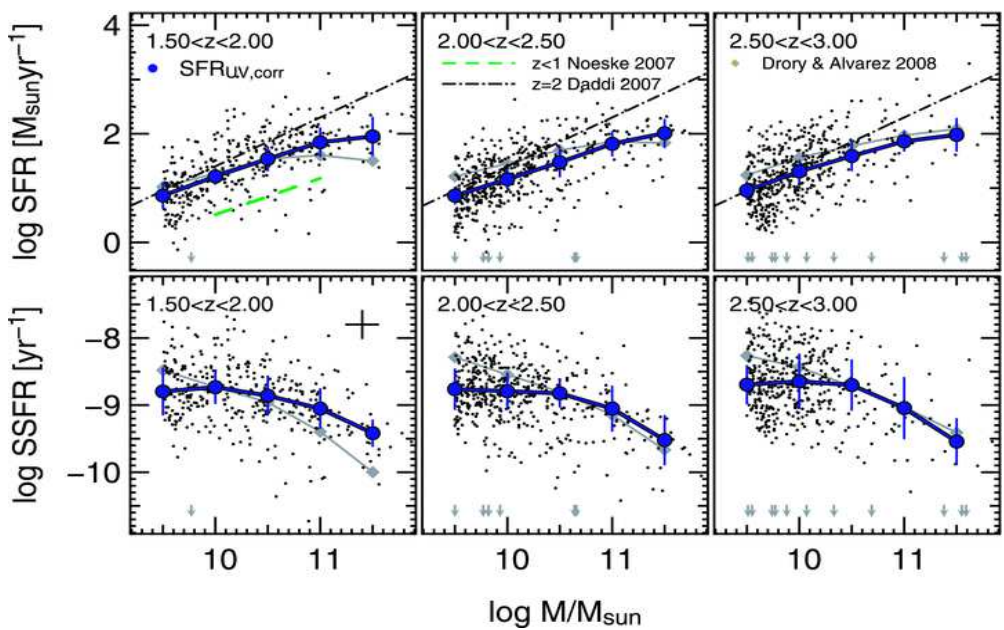} 
 \caption{Relation between stellar mass and star formation rate for galaxies
from log $M_{*} = 9.5 - 12$. The star formation rate increases with 
stellar mass, on average, during this epoch although the specific star
formation rate drops at higher masses.  The star formation relation 
with mass is also fairly constant during this epoch (from Bauer et al. 2011). }
   \label{fig1}
\end{center}
\end{figure}

There is a strong signature of galaxy `downsizing' in almost
all galaxy properties up to at least $z = 3$, such that the highest 
mass galaxies tend to `shut down' formation modes, including AGN, 
before the lower mass galaxies.  Essentially this means that high
mass galaxies finish forming before lower mass ones.   
For example, when examining the 
abundances of massive galaxies with  M$_{*} > 10^{11}$ \solm, the 
measured number densities 
up to $z \sim 1-2$ are similar to what is found in the local universe, at
the same stellar mass limit (e.g., Conselice et al. 2011b; Mortlock et al. 2011).
Therefore to study the properties of the the most massive galaxies and to
examine their formation observationally we must go to higher redshifts, 
namely at $z > 2$.  

Recent results have accomplish this by examining the most massive systems 
at 1.5 < z < 3 with Hubble Space Telescope surveys that can resolve
these systems.  This includes the CANDELS survey (Grogin et al. 2011) and
the GOODS NICMOS Survey (GNS; Conselice et al. 2011b).  Recent results from 
these surveys have shown that the formation modes for the most massive 
galaxies are dominated by mergers (Bluck et al. 2009; 2012), and the 
accretion of gas from the intergalactic medium (Conselice et al. 2012) with
the bulk of this formation occurring at $z > 1$ (Mortlock et al. 2011).  We
explain below the reasoning behind these results, and how others have found 
similar conclusions.  The below analysis has only been carried out for
the highest mass galaxies so far, although in the future lower mass
systems can be analysed in a similar way using deeper data.

\section{Empirical Galaxy Formation}

\subsection{The Role of Mergers up to $z = 3$}

Galaxy assembly is a combination of at least three processes. These are:
merging with existing galaxies; the accretion of cold gas from the
intergalactic medium; and the conversion of in-situ initial gas into
stars in a galaxy over time. Understanding the relative role of these 
processes, and how these vary as a function of stellar mass and 
environment, is one of the major goals of extragalactic astronomy.

We now have some idea about the role of mergers in galaxy assembly (e.g.,
Conselice et al. 2003; Lotz et al. 2008; Bluck et al. 2012).  While mergers can dominate much
of the evolution within galaxies, including triggering star formation and AGN,
instigating morphological changes, etc., we are mainly interested here in how
it builds up the stellar masses of galaxies over time.  

The amount of stellar mass added to a galaxy due to the merger
process is given by the integral over the merger history, based on the
fraction of galaxies merging, and the time-scale for mergers (e.g.,
 Bluck et al. 2009; 2012).  Bluck et al. (2012) and Conselice
et al. (2012) carry out this integration 
using the observed merger history measured
directly from the GNS sample of massive galaxies at $ 1.5 < z < 3$ (Fig. 4), 
and the modeled time-scale for mergers (e.g., Bluck et al. 2009; 2012).
 
Using pairs of galaxies and galaxies involved in merging, as seen 
through the CAS
system, we can now measure accurately the merger history up to $z = 3$ (e.g.,
Bluck et al. 2012 and references within).     The total amount of stellar 
mass accreted into a galaxy is a double integral over the 
redshift range 
of interest ($z_1$ to $z_2$ corresponding to look-back times $t_1$ and $t_2$), 
and over the stellar masses which we probe
($M_1$ to $M_2$), which for 
the GNS, sensitive down to M$_{*} = 10^{9.5}$ \solm, can be expressed as,

\begin{equation}
{\rm M_{*, M}} = \int_{t_{1}}^{t_{2}} \int_{M_{1}}^{M_{2}} M_{*} \times \frac{f'_{m}(z,M_{*})}{\tau_{\rm m}(M_{*})} dM_{*} dz,
\end{equation}

\noindent where $\tau_{\rm m}$(M$_{*}$) is the merger time-scale, which 
depends on the
stellar mass of the merging pair (Bluck et al. 2012).  The total 
integration of the amount of mass assembled through merging gives 
${\rm M}_{\rm *,M}/{\rm M_{*}(0)} = 0.56\pm0.15$, where M$_{*}(0)$ is 
the initial average stellar mass of the GNS massive galaxy 
sample.  This is the fractional amount of stellar mass added due to both 
major and minor mergers
for systems with stellar mass ratios down to 1:100 for the average massive
GNS galaxy after following a merger adjusted constant co-moving 
density (Conselice et al. 2012).

However, to fully understand the total baryonic mass assembly of galaxies 
due to the merger process, we also need to account for how much gas mass is 
brought into these systems through mergers.  This is calculated by 
integrating the amount of gas in
these merging systems using an empirical fit to the relationship
between the gas mass fraction, $\mu_{\rm gas}$, and the stellar mass found
at $z = 2-3$.  Overall the lower mass galaxies contribute the bulk of the 
gaseous mass from mergers, whereas most of the stellar mass accreted in 
mergers 
arises from higher mass ratio mergers.  We show this relative role of mergers 
in Figure~4.     This relation can then be used to calculate for
the GNS sample how much gas mass is added due to merging, finding 
M$_{\rm g, M}$/M$_{*}(0)$ = M$_{\rm *,M}$/M$_{*}(0)$ 
$\times$ f$_{\rm g}$ = 0.57$\pm0.15$.  Over the redshift interval
$z = 0 - 3$ major and minor mergers are roughly equal in terms
of importance in building up galaxies.  

The high number of minor mergers is also a solution to the size problem in
massive galaxies.  When examining the effective radii of massive galaxies at
redshifts $z > 0.5$ they are much smaller than galaxies of similar
masses in today's universe (e.g., Buitrago et al. 2008; Trujillo
et al. 2007; Weinzirl et al. 2011).   How these galaxies expand to become the large
galaxies we see today is not well understood.  However, by examining
the number of minor mergers we observe from $z = 3$ down,
and by using simple physics, it is possible to show that these
mergers provide enough mass at the right locations to expand the
measured sizes of galaxies by up to a factor of five (Bluck et al.
2012).  This is therefore likely the solution to the problem of
how galaxies can be so compact at high redshift, but still have
significantly high stellar masses (see also Ownsworth et al. 2012).

\begin{figure}[b]
 \vspace*{-0.3 cm}
\begin{center}
 \includegraphics[width=5.5in]{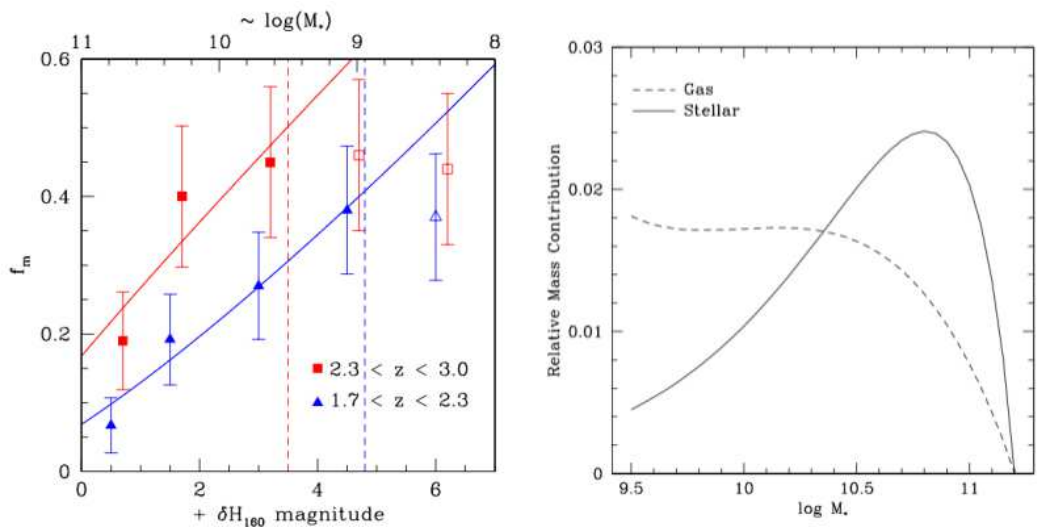} 
 \caption{Plots showing the role of mergers, both major and minor, and gas accretion in forming galaxies at $z = 1-3$.  The left panel shows the merger fraction for systems with stellar masses log M$_{*} > 11$.  Shown at the top of the left panel are the stellar masses of the lower mass galaxies merging with these massive galaxies down to log M$_{*} = 9.5$, demonstrating that minor mergers are more common than major mergers (Bluck et al. 2012).  The right panel shows the relative amounts of mass added to these massive galaxies due to mergers, revealing that most of the gas is brought in by lower mass galaxies while higher mass galaxies add most of the stellar mass (Conselice 
et al. 2012.) }
   \label{fig1}
\end{center}
\end{figure}

\subsection{Gas Accretion From the Intergalactic Medium}

One important observation of high redshift massive galaxies with log M$_{*} >
11$ \solm is that these systems have an average star formation rate that is
relatively constant at $1.5 < z < 3$, and declines at $z < 1.5$ (Fig. 3).
This star formation increases the stellar mass within these systems
by an amount which approximately doubles it.
This is a high star formation rate, and the amount of gas
mass accreted due to merging, plus the original amount of gas is not enough to
sustain the star formation present (Papovich et al. 2011; 
Conselice et al. 2012).

We can show, based on this, that the amount of gas accreted
into a massive GNS galaxies is 
${\rm M}_{\rm g, A}/{\rm M}_{*}(0) = 0.70 \pm 0.22$ (Conselice et al. 2012), such that an amount of
mass on the order
the entire initial stellar mass of a massive galaxy is added
over time outside of mergers to form stars during $1.5 < z < 3$, a time
span of $\sim$ 2 Gyr.  This reveals a net gas accretion, which is then turned
into stars, of  $61 \pm 19$\,\solm ${\rm yr}^{-1}.$
When considering that these galaxies have outflows 
(e.g., Weiner et al. 2009) that could easily double the amount of gas mass
needed to be accreted from the IGM (e.g., Faucher-Giguere et al. 2011)
we find that the gross inflow rate increasing to: 
${\rm \dot{M}_{\rm acc}}= 96\pm19\, {\rm M}_{\odot}\, {\rm yr}^{-1}. $

The result of this is that gas accretion accounts for 49$\pm$20\% of the 
stellar matter added to galaxies from $1.5 < z < 3$.   Mergers account 
for the remainder of the mass assembly, with 1/2 of this minor mergers 
and 1/2 of this major mergers (Bluck et al. 2012). Gas accretion however 
is responsible for 66$\pm20$\% of all new star formation during this epoch within
log M$_{*} > 11$ galaxies. 
Overall this implies that gas accretion into massive galaxies
at early epochs is potentially a major formation method, and dominates over 
mergers as a formation mechanism for new stars.  This is however
a first estimate of this quantity, and
future studies with wider and deeper surveys will measure this number
with more accuracy in the future. 

This measured gas accretion rate is roughly consistent with theoretical 
calculations which predict a similar amount of gas accretion 
(e.g., Murali et al. 2002; Dekel et al. 2009).   Some of the 
first predictions of the gas accretion by Murali et al. (2002) 
found a gas accretion rate of $\dot{M}_{\rm g, A}$ $\sim 40$ \solm yr$^{-1}$,
while more recent work suggests higher rates of 
$\dot{M}_{\rm g, A}$ $\sim 100$ \solm yr$^{-1}$
(e.g., Dekel et al. 2009; Faucher-Giguere et al. 2011). These first results
on measuring cold gas accretion are in
general agreement with these models, although this measurement should be
redone for larger samples and at lower masses to determine the role of mergers
vs. accretion as a function of stellar mass.

\section{Role of Environment}

\subsection{Galaxy Clustering and Dark Matter Halos}

One of the new frontiers of studying galaxies and their evolution is to
examine how they cluster together, and what this reveals of their halo
masses, as well as their environment.  The summary of how this evolution occurs
is shown in Figure~5 based on results from Foucaud et al. (2010).  
In general the most massive
galaxies are the most clustered, with correlation lengths of 
r$_{0}=$ 10-15 h$^{-1}$ Mpc from
redshifts $0.5 < z < 2$ (Hartley et al. 2010). Interestingly, there 
does not appear 
to be a strong trend with redshift, such that
the most stellar massive systems have the highest clustering values up
to $z = 2$.  At higher redshifts, the clustering strength increases at a fixed
stellar mass, such that the most massive halos assemble their stellar mass
before lower mass ones.

The clustering strengths of these galaxies can be converted into halo masses
by matching the abundances of galaxies selected by stellar mass into a
corresponding halo mass from N-body simulations within a given 
cosmology (e.g., Mo \& White 2002).   Using this method it is therefore possible to determine
how the stellar to total mass ratio evolves for ensembles of galaxies as
selected through their stellar mass.  When comparing total masses to stellar
masses the most massive systems have the lowest stellar to total mass ratios
(Fig. 5).  This implies that something is either shutting off star formation
within the most massive galaxies, or more likely that halo masses contain
sub-halos which contribute to their total masses.   Systems
with total masses M$_{\rm halo} > 10^{13}$ \solm have stellar mass to halo mass
ratios of $< 0.01$ while this ratio goes down to $\sim 0.1$ for systems
with M$_{\rm halo} \sim 10^{11}$ \solm, although we know that dwarf
systems also have very low M$_{*}$/M$_{\rm halo}$ ratio (Penny
et al. 2009). 

Related to the stronger clustering at higher redshifts, it is also clear
that at fixed stellar mass, the ratio of stellar to halo mass increases
at lower redshifts at a given halo mass.  This implies that there is a halo downsizing, such
that the most massive halos form stars first, and only after 
a few Gyr do lower mass halos form enough stars for their systems to
enter the same stellar mass selection. This thereby lowers the ratio of
stellar to halo mass,  given that these late comers to a given 
mass bin are in lower mass halos, which lowers the clustering
strength of the population selected by this stellar mass range.  Note
however, that is it not clear if this is a general result or simply
for galaxies selected by stellar mass.  For example
disk galaxies up to $z = 1.4$ shows very little evolution in the
stellar mass to total mass ratios (Conselice et al. 2005b), or
alternatively in the stellar mass Tully-Fisher relation (e.g., Conselice
et al. 2005b; Miller et al. 2011).   

Integral field spectroscopy for $z > 1$ systems has
also revealed important clues about the nature of these high 
redshift galaxies (F\"orster Schreiber et al. 2009).
These studies typically find an equal distribution of systems
which are: 1. rotationally dominated, 2. mergers and 3. 
systems that have high velocity dispersions, but which are
very compact.  The true nature of these systems has yet to
be revealed, but clearly the kinematics of distant galaxies
shows significant differences from galaxies in the local
universe.  More IFU studies of larger samples of galaxies
are currently needed, as there are at most only a few hundred
IFU spectra measured for $z > 1$ galaxies thus far.

\begin{figure}[b]
\begin{center}
\hspace*{-0.5cm} \includegraphics[width=6.5in]{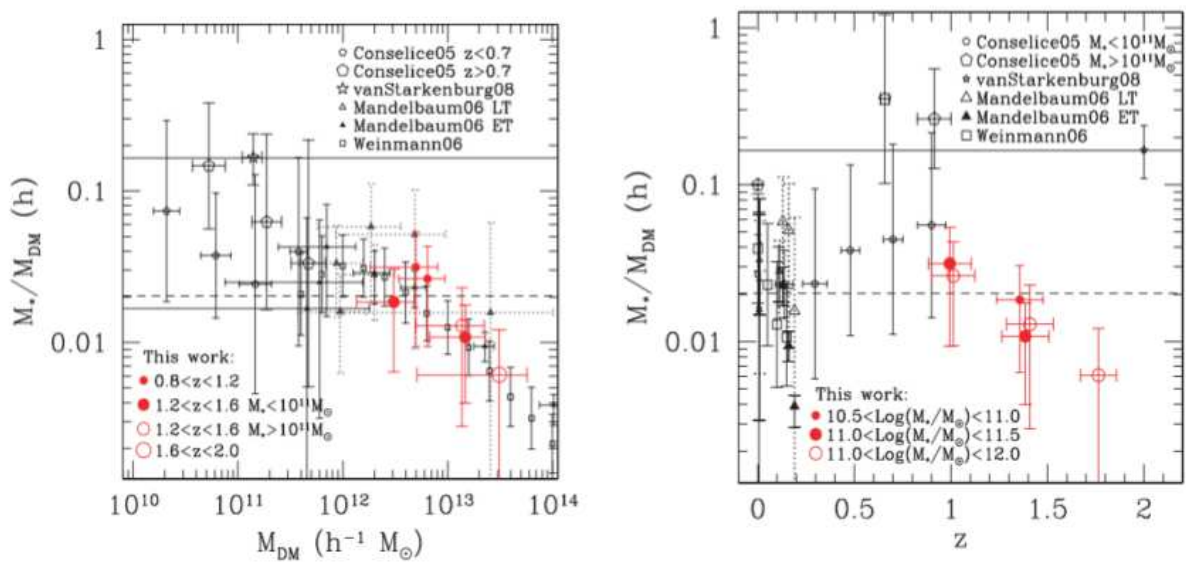} 
 \caption{The relationship between the halo mass and the ratio of stellar to halo mass (left panel), as well as this ratio of 
masses as a function of redshift (right panel) (Foucaud et al. 2010).   As shown, the amount of stellar mass relative
to dark matter mass declines at higher redshifts for systems with larger halo masses.   Furthermore, there is some evidence
that the highest mass halos are populated earlier than low mass ones, with the result being that for a given
stellar mass  selection there is a higher ratio of stellar to dark matter mass at lower redshifts.   }
   \label{fig1}
\end{center}
\end{figure}

\subsection{Environment vs. Mass - Which Dominates?}

Since galaxies were first studied, and especially since the paper
by Dressler (1980) there has been a recurring question of the role
of environment in driving galaxy formation/evolution. It is clear
that in the local universe disk galaxies are more likely 
found in low density environments, while early-type systems are
found in denser ones (e.g., Dressler 1980).  However, this effect
is most pronounced in extreme environments, and it is not
clear beyond these extremely dense environments how, say a modestly
dense environment would affect galaxy evolution over a very low
density environment.

This was examined in detail by Gr\"utzbauch et al. (2011a,b)
who looked at galaxy colours and star formation rates as a function
of both environmental density as well as a function of stellar mass and
redshift.
What was very clear is that while there is some environmental
effect, such that galaxies are redder in denser areas (Gr\"uzbauch
et al. 2011a), this effect is most pronounced at lower redshifts,
typically at $z < 1$.   At redshifts higher than this the
effects of environment are very minimal (e.g., Gr\"uzbauch
et al. 2011a).

\begin{figure}[t]
\begin{center}
\hspace*{-0.5cm} \includegraphics[width=5.5in]{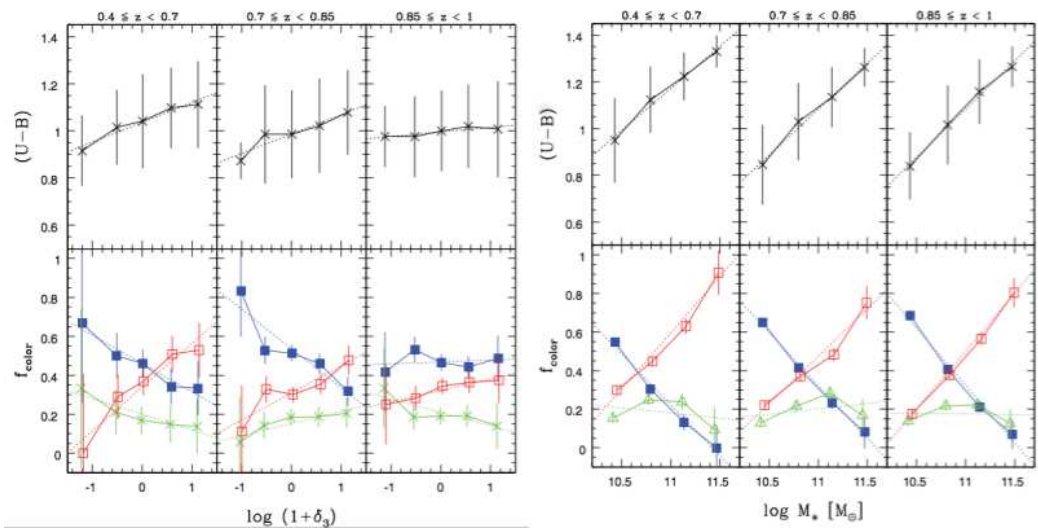} 
 \caption{The right panel and left panels show the relationship between
galaxy rest-frame colour and environment (as measured through 1+$\sigma$)
and stellar mass (Gr\"utzbauch et al. 2011a).  This relations are also shown as a function of redshift
in three panels.  The left panels shows how the relationship between
environment and colour such that there is very little trend between these,
while the right panel shows that there is a strong correlation between
the stellar mass and colour, such that higher mass galaxies are always
on average redder than lower mass systems. This strong trend continues up to
$z - 3$ (e.g., Gr\"utzbauch et al. 2011b). }
   \label{fig1}
\end{center}
\end{figure}

On the flip-side of this, when comparing galaxy colours and
star formation rates with stellar masses, there is a much
stronger correlation, such that the most galaxies have
red colours compared to lower mass galaxies, as well
as having a higher star formation rate at higher redshift
(Gr\"utzbauch et al. 2011b; Fig. 6).    There is also very little
trend in galaxy properties with the overall halo mass in
which a galaxy is located, demonstrating that environment,
as measured by the number of nearby galaxies and the
total mass of the group/cluster a galaxy is located
has very little effect on the observed properties of
galaxies.  This is consistent with there
being very little trend in age or star formation history
for local galaxies as measured through stellar spectra
fitting (e.g., Thomas et al. 2005; \S 2.1).

These correlations show that the stellar mass, or likely
the halo mass of an individual galaxy is the overall most
important aspect for how these systems form and evolve.  
This is possibly related to the fact that galaxies
with higher masses have more massive black holes, and
that AGNs are significantly active during this epoch,
depositing 35 times the binding energy into the galaxy
over $1 < z < 3$ on average
(e.g., Bluck et al. 2011).   Larger black holes
produce more energy back into their host galaxies,
and this is possibly the reason why mass is
such as defining characteristic of galaxies.  This may
also explain how downsizing occurs first for the most
massive galaxies, although finding direct proof that
AGN/black holes have a significant effect on gas
and star formation in galaxies remains elusive.

\section{How Well Does Theory Predict Galaxy Evolution?}

Understanding how galaxy formation occurs was initially first calculated using the ages of 
different stellar populations in the Galaxy (Eggen et al. 1965), and the 
default initial assumption was that galaxies formed
like stars in a monolithic-type collapse.  In the 1980s the first 
computer simulations 
of structure formation  showed 
that a universe dominated by Cold Dark Matter (CDM) matched observations of galaxy 
clustering on large scales (Davis et al. 1985), and that within this 
framework galaxy  assembly should be hierarchical (Blumenthal et al. 1984).

The situation today is that there are many simulations that are used to 
predict properties of the galaxy population, and how it evolves through
time.  Significant success has been reached when predicting the properties 
and scaling relationships of galaxies, yet problems still exist
(e.g., Guo et al. 2011).  While there are famous problems such as the 
satellite and the CDM dark matter profile problem, there are also
issues when examining 
how the evolution of galaxies occurs, and trying to match this with the theory.

\begin{figure}[b]
\begin{center}
\hspace*{-0.5cm} \includegraphics[width=5.5in]{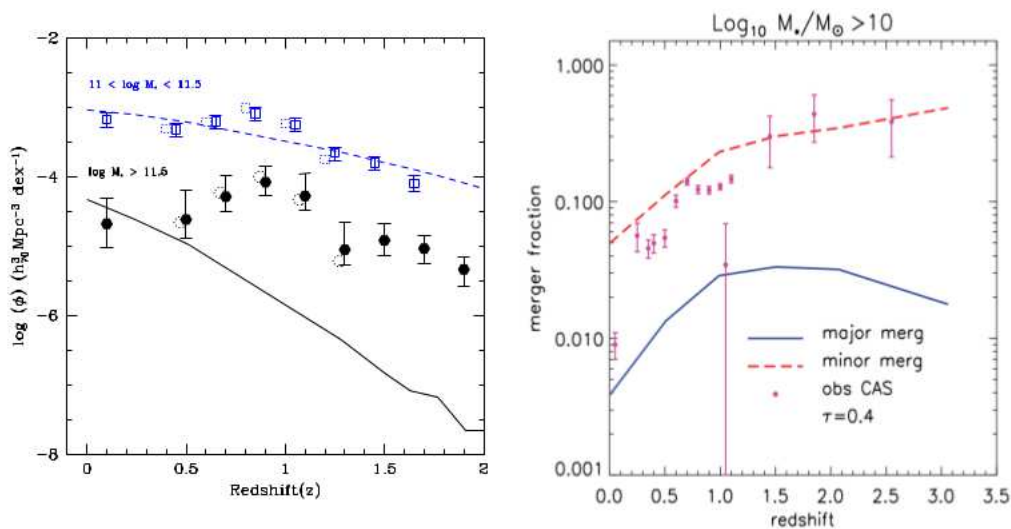} 
 \caption{Plot showing issues with CDM simulation results.  The left panel
show the evolution of the number densities for galaxies with stellar masses between
M$_{*} = 10^{11} - 10^{11.5}$ \solm and galaxies with M$_{*} > 10^{11.5}$ \solm.
For the most massive systems there is a significant difference in the numbers of
massive galaxies and the CDM prediction from the semi-analytical Millennium 
simulation, up to a factor of 100 (Conselice et al. 2007).  The right panel shows the comparison between
the major merger history and simulations based on CDM (Bertone \& Conselice 2009).  
The blue solid line shows the prediction for
the same quantities as the points, demonstrating values that are significantly lower
than the observations. }
   \label{fig1}
\end{center}
\end{figure}

I focus here on the much less well known problems of CDM in predicting galaxy 
evolution at high-redshift.  While semi-analytical CDM models can predict
local galaxy properties well (e.g., Bower et al. 2006; 
Guo et al. 2011), there are very significant
differences between observations and theory when probing at higher redshift.
One of these is that most semi-analytical simulations are not able to 
reproduce 
the abundances, or the formation history of massive galaxies through mergers 
(e.g., Conselice et al. 2007; Bertone \& Conselice 2009; Marchesini et al. 
2010; Guo et al. 2011; for further information on problems in other
galaxy predictions using CDM see e.g., Guo et al. 2011).

Specifically, the number densities of  massive galaxies with 
log M$_{*} > 11$ are 
under-predicted in CDM galaxy formation models at $z > 2$ 
(e.g., Conselice et al. 2007; Marchesini et al. 2010; 
Guo et al. 2011; Fig. 7).    The difference with CDM models can be
up to a factor of 10 or higher up to $z = 2$.  While some CDM models
attempt to get around the star formation downsizing through merging existing,
but quiescent, galaxies at $z < 1$ (De Lucia et al. 2006), these systems are 
clearly already well formed by this time. To further investigate the
 problem of matching observables with theory requires 
that we investigate how the formation process of 
galaxies occurs, and whether models can reproduce these known
formation modes.  One of major 
methods for doing this is to investigate how well CDM models can reproduce 
the formation history of galaxies as seen through processes such as merging.   

For example, Bertone \& Conselice (2009) compare the merger history of 
galaxies to the predictions from the Millennium simulation.  This 
comparison shows that the Millennium simulation underpredicts the number of 
major mergers by a similar order of magnitude (factor of 10) that it 
underpredicts the abundances of galaxies (Fig. 7).  The reasons for this are 
unclear, but may relate to either underlying cosmological 
assumptions, or the way in which baryons are implement in these simulations.  
Future work, including investigating the underlying role of dark matter, and 
other cosmological features, will have to be included in future theoretical
research on how galaxies form.

\section{Summary and Outlook for the Future}

The basic ideas presented in this review is that we can observationally
determine how galaxy formation occurs, and do not have to rely on
comparing basic observables to models to understand this history.  I
have shown how we now have some understanding for how the most
massive galaxies with M$_{*} > 10^{11}$ \solm
assembled their stellar mass and baryons at $z < 3$ -- showing that
likely mergers and gas accretion are equally important during this
epoch.  Observations of higher redshift galaxies are still very
new, especially in terms of having stellar mass completed samples,
although progress will be quickly made with surveys such as CANDELS,
UKIDSS UDS and VISTA.

While we have learned quite a bit about the history of the formation of
massive galaxies, probing the star formation and merger rates, there is still 
a significant amount regarding galaxies that we are still just starting
to explore.  Some of these outstanding issues that have not been fully 
addressed, 
include: how disk galaxies and their structures (spiral arms, bars) 
form, and what the nature of massive galaxies themselves are at 
$z > 2$, as these 
systems are not analogs to any local galaxies in almost every way. 

Deep and wide field surveys such as with Euclid and WFIRST will carry out large 
imaging surveys
that will address the problem of galaxy evolution in great detail, providing
the large and deep fields that Hubble and ground based telescopes cannot
provide.  However, what is also needed is kinematic measurements alongside 
structural measurements to truly decipher the nature of distant
galaxies.   We are just starting to scratch the surface of what can
be done, but in the near future instruments such as KMOS on
the VLT, and in the
future the ELTs will provide large numbers of IFU spectra through
surveys.  

We have also not yet detected the first galaxies to form, although we
are pushing the limits (e.g., Yan et al. 2012).  JWST and the ELTs will
provide a deep probe of the universe such that we will likely see the
first galaxies and perhaps stars forming, especially utilising the
benefits of gravitational lensing (e.g., Coe et al. 2012).  
The next generation of radio telescopes, such as SKA and its 
precursors will furthermore allow us to measure the still largely
uncertain properties of gas within distant galaxies, including the
important question of how the gas content evolves with time.  This
will allow us to complete our physical and empirical knowledge of
galaxy formation.

\vspace{0.5cm}

I thank my collaborators, students and post-docs for allowing me to include 
some of their work in this review, and for the numerous discussions that
have significantly increased my understanding of the topic of galaxy formation.  This
includes: Alice Mortlock, Asa Bluck, Fernando Buitrago, Matt Hilton, Ignacio Trujillo,
Shardha Jogee, Tim Weinzirl, Will Hartley, Jamie Ownsworth, Ruth Gr\"utzbauch, Seb
Foucaud,
Omar Almaini, Amanda Bauer and Ken Duncan.  Support for some of the
research presented here came from STFC, NASA, NSF and the Leverhulme Trust.

\end{document}